# Reflexivity and the diagonal argument in proofs of limitative theorems

Kajetan Młynarski
kajtek@iphils.uj.edu.pl

Institute of Philosophy, Jagiellonian University
Kraków, Poland

October 29, 2018

## Abstract

This paper discusses limitations of reflexive and diagonal arguments as methods of proof of limitative theorems (e.g. Gödel's theorem on *Entscheidungsproblem*, Turing's halting problem or Chaitin-Gödel's theorem). The fact, that a formal system contains a sentence, which introduces reflexitivity, does not imply, that the same system does not contain a sentence or a proof procedure which solves this problem. Second basic method of proof - diagonal argument (i.e. showing non-eqiunumerosity of a program set with the set of real numbers) does not exclude existance of a single program, capable of computing all real numbers. In this work, we suggest an algorithm generating real numbers (arbitrary, infinite in the limit, binary strings), and we speculate it's meaning for theoretical computer science.

## 1 The Problem

Proofs of basic limitative theorems such as Gödel's theorem [Goe31, Goe86], Turing's theorem about undecidability of the halting problem [Tur36] or Chaitin-Gödel theorem [Cha74, Cha02], are proofs conducted by contradiction, through application of two kinds of reasoning:

- diagonal argument
- arguments based on reflexivity

A typical example of the latter is the Liar paradox (Gödl, Turing) or Berry's paradox (Chaitin). Furthermore, part of these proofs are conducted through contradiction. Both ways of reasoning, (reflexivity in particular) are treated with a limited trust as *sui generis* "tricks" or results, which do not have a lot in common with a real computational practice [Heh] [1].

## 2  Arguments from reflexivity

Let us denote the sentence "This sentence is false" by $A$ [2] and "the sentence on the left of this conjuction is true" as $ap$. Then each sentence of the form: $A \wedge ap$ is always false, because if the antynomy is true it states false, and if it states false it is true. At the same time this conjunction is very natural for "common" non reflexive sentences, and it has the same logical value as they do.

Let us define a formal system with the following axiom $(AK)$:

$$\forall z \in Z, W(z) = W(z \wedge ap) \tag{1}$$

where $Z$ is a set of sentences of arbitrary complexity and $W(z)$ is a logical value of sentence $z$. In such a system, we define value of a sentence as its conjuction with $ap$. In particular, we define in such a way, values of all reflexive formulas (e.g. multi-proposition antynomies). One can suppose, that such system should be free of problems corresponding to semantic reflexivity.

If we want to free the system of limitation determined by Gödel's theorem we meet a certain complication. Gödel syntactically described a semantic paradox, namely the reflexive sentence: "*T*his sentence is non-provable in this system". That is why it has a rank of a theorem. One should also notice, that alternative in the form $A \vee ap$ is a true sentence, according to the rules of propositional calculus. It is, however explicitly inconsistent with basic semantic intuition.

## 3  The halting problem

The sentence $ap$ is clearly semantical. Therefore to obtain proper "delimitative" theorem, one should describe it syntactically. Let us try for the "computational equivalent" of Gödels theorem, namely for the theorem about undecidability of the halting problem

The proof of this theorem is performed with Russel's method i.e. by contradiction and application of the reflexivity property:

- Let us assume, that there exists program $T$ (tester), which for arbitrary program $P$ and input $D$ returns 1 if $P$ stops and 0 if $P$ runs infinitely. Therefore $\forall P, D T(D,P) = 0\, T(D,P) = 1$.

---

[1] Hilbert continued works on formalization of mathematics, despite Gödel's results, which strongly impressed philosophers, rather than mathematicians. Zermelo did not care about those theorems as well

[2] One should note, that from the computer science point of view, sentence formulating the Liar paradox is presumably the shortest known "infinite-loop" algorithm (it does not produce any output). Formally, in turn, it is the shortest non-overflow system, which includes one sentence and one axiom used to proove it.

- We construct program $S$, which runs in endless loop if $T$ returns 1 and it ends if $T$ returns 0

- If we provide $S$ as an input of $T$, then $T$ goes into an infinite loop and does not decide about the halting problem.

Therefore, against assumptions, the universal tester does not exist, and the halting problem is in general case undecidabile.

Let us now construct a "large tester" **T** . It consists of a "common" tester $T$, as in the previous proof of halting's problem undecidability, and an additional tester $T'$ of the following properties:

- If $T$ does not run in an infinite loop, then $T'$ runs in an infinite one.

- If $T$ runs in an infinite loop, then $T'$ returns 0

- $T'$ is basically very similiar to $T$ (difference is in the output states)

- $T'$ tests T only, when running parallel with it.

Proof of the delimitative theorem:

- Let us assume, that there exists program $T$ and it is a universal tester determining the halting problem.

- Let us assume, that there exists program $S$ running in an infinite loop if the value of $T$ equals to 1. In such case t will loop infinitely and $T'(T = 1, S) = 0$ and such value will be returned by $T$, which does stop. Therefore thesis about existence of $T$ is not contradictory with the assumption. Let us observe also, that no program provided on input, can not cause $T$ to run in an infinite loop, since $T'$ examines only the state of $T$.

## 3.1 Remarks

Both programs work in paralell, that is they process their input at the same time.

Hypothesis: **T** does not work if $T$ and $T'$ do not run in parallel. This hypothesis has a meaning for the problem of equivalence of parallel and sequential processing, and also for the understanding of semantic structure processing ("The secret of semantics is hidden in parallelity").

The fact, that **T** can solve the halting problem, does not mean that it is going to do it in finite time, because the tested program can (as an object) have infinite, irreducible computational complexity, or use random generator and never loop in a regular pattern.**T** will, in such a case, run the test infinitely long. Lack of decision is caused, however, by a property of the tested program (input data) and not insufficiency of **T** . One can extend **T** with an additional algorithm, which gives an output in regular time intervals during the computation. Such an algorithm allows empirical testing of the halting problem with arbitrary (finite) precision. There is also possibility of constructing an algorithm, which detects "random loops" and prints warning about "possible lack of halt". In summary: such a tester can work infinitely long, processing infinite or incomplete data.

Let us assume, that there exists a formal system in which each true sentence can be proven. In such a system, sentence: "this sentence is unprovable" is equal to sentence: "this sentence is false". Therefore sentence used in Gödels proof becomes Liar's antynomy. If the system contains axioms of $AK$ type, it becomes a possible system (with respect to Gödel's limitation).

Now, let us assume existance of a formal system in which one can proove every true sentence from some field e.g. the number theory. Sentence "this sentence is unprovable" means then, that it does not apply to this theory. Indeed the sentences, which for instance, refer to the yearly rainfall on Sahara do fulfill this requirement. In turn, the sentence "this sentence refers to the number theory and it is unprovable" is a "local" Liar's antynomy. In such a case, conjuction: "this sentence refers to number theory and it is unprovable $\wedge$ the sentence on the left side of this conjunction is true with respect to the number theory" is a false conjuction. Because of that, the assumed system, can in principle exist, provided that it contains an axiom analogous to $AK$.

The role of confirming sentences of $ap$ type, shows the great importance of the truth decision, challenged by deflationistic truth theoreticians.

In proofs of Gödel's theorems, the diagonal argument plays an important role and it should be also considered.

## 4 The diagonal argument

In case of the diagonal argument, we will consider theorem about existance of real, non-computable numbers proven by Alonso Church [Chu36b, Chu36a]. It states the following: "*t*here exists uncountably many natural numbers, which are non-computable by finite algorithms ". In the simplest way the proof is done by showing that:

- Each real number can be expressed as an infinite binary string

- There are $2^{\aleph_0}$ infinite binary strings

- Each finite algorithm could be expressed by a finite binary string of arbitrary length

- There are $\aleph_0$ binary strings of arbitrary length

- Therefore each computable number corresponds to infinitely many uncountable numbers (i.e. coded with infinite, non-compressible strings)

This proof is done by contradiction without referring explicitelly to the diagonal argument, however meaning of this argument relies on showing non-equinumerosity. The fact, that there exists less finite algorithms than real num-bers, does not imply, however, that the finite algorithm allowing to compute all real numbers with an infinite precision does not exist.

**Theorem 1.** *There exists a finite algorithm capable of computing all real numbers (and or binary strings) with an arbitrary precision (exactly in limit).*

*Proof.* It is a simple algorithm (AS) generating all arbitrary long binary strings, therefore allowing computation of all real numbers with an arbitrary precision. It operates on a binary tree, as shown in figure 1: □

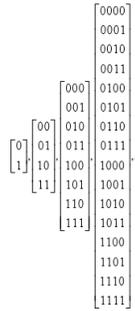

Figure 1: Construction of a binary tree

In its $n-th$ step $AS$ generates all strings of length $n$ (the diagonal argument does not apply because obtained matrices are rectangular and complete). If we decide to perform a limit transition, we will obtain a matrix of size $\aleph_0 \times C$. One can easily observe, that the set of real numbers (infinite binary strings) obtained in such a way is ordered by generating process, which is non-continous and non-dense. One can, in principle generate in the limit, sets of higher power e.g. by requiring $2^{2^n}$ growth. It is, therefore, possible to generate (in the limit) a set equinumerous with the set of natural numbers, through adding one string in each step:

In each step we obtain a square matrix. By applying the diagonal argument, we show, that certain arguments were ommited. Indeed, this algorithm generates only $2^{\log_2(n)}$ strings, out of $2^n$ possible.

**Definition 1.** *A real number or binary string is called a well defined if its definition contatins full information, unequivocally identyfying this number. Such definition can, of course, be itself an infinite string.*

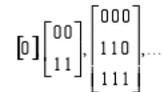

Figure 2: Construction of square matrices.

**Theorem 2.** *There exists a finite algorithm, allowing for a computation of every well defined number and/or binary string with an arbitrary precision (exactly in the limit).*

*Proof.* We construct an algorithm $AS + C$ consiting of $AS$ and algorithm $C$ which sequentially compares obtained strings with a defining number (string). If a number (string) is well defined (full information), then this task is possible. Therefore it is possible to obtain any well defined number with an arbitrary precision. □

Unfortunatelly, precise definition may be infinite, hence unreachable.

**Theorem 3.** *There exists an algorithm allowing for exact computation of single, arbitrary real numbers / binary strings*

*Proof.* It is algorithm $AS$ enriched with algorithm $L$, which randomly chooses one path on a binary tree. Consequitve strings are more and more precise expansions of a certain real number. Since generation of a binary tree is a hard problem(requiring exponential time) this procedure may be simplified, for instance in a following way: with each cycle $ASL$ chooses only one string, and applies $AS$ to it. □

Requirement for truth of this statement is existance of real random number generators. Pseudorandom generators are coded with finite binary strings, therefore they generate numbers which are computable and compressible. Algorithm $ASL$ does not differ in a substantial way from "normal" generation of a binary string, however its structure may be useful in proving following theorems.

**Definition 2.** *We call a binary string compressible, if there exists an equivocal relation assigning a shorter string to it.*

Let us observe:

- There are $2^n$ strings of length $n$
- There are $\sum_{i=1}^{n-1} 2^i = 2^n 2$ shorter strings
- Therefore half of the strings can be compressed with one bit, one fourth with two bits etc.
- Therefore, except at most two, all binary strings of a finite length are compressible. However, compression of just a few bits is practically not interesting.The possibility of compressing most of strings is therefore negligible.

**Definition 3.** *A binary string of length n is called m-non-compressible if there existis no program encoded by a string of length m+c (where c denotes length of a compresing program without data, which )*

One can imagine a "compression machine", which assigns shorter strings (names) to longer strings. A requirement for performing compression, is being in "posession" of an appropriate number of names. Such a machine is useful to inspect general limits and properties of compressibility. It is clear, that two different machines can compress different strings with non-equal efficiency. For instance: $M1$ can compress a string 001001110110100010101001111111001 to a string 1, and $M2$ can not compress it at all (name identical with compressible string). Therefore, for some problems some machines will be more useful than other. Anyway, if we could use many machines, we could improve compression. One has to see, that such machines have to work independently (in other way, they could be replaced by a single machine with necessary constraints). It is possible, that our brains operate according to those principles.

**Theorem 4.** *There exists $2^n - 2 - \sum_{i=1}^{i=m} 2^i$ m-non-compressible binary strings of length n.*

Proof is trivial, and results from the definition of compressibility.

**Theorem 5.** *There exists an algorithm, allowing for arbitrary precise computation of single real numbers "uncomputable", and/or binary strings which are only m-non-compressible*

*Proof.*
- For each assumed method of compression, it is possible to construct a filtering algorithm $E$, which rejects all m-compressible strings from the list of $n$-long strings generated by $AS$

- Let the algorithm $L + E$ generate a random path, allowing only for $m$-non-compressible strings. Complex algorithm $AS + L + E$ computes m-noncomputable number or m-noncompressible binary string. In such a way, we obtain a binary string, which is "guaranteed" to be non-compressible or a "noncomputable" real number, "guaranteed" to be noncomputable. □

Therefore it is possible to compute also single uncomputable numbers, knowing that obtained strings represent exactly those numbers.

**Theorem 6.** *There exists an algorithm which allows to decide, whether a given string is m-non-compressible, which can be written with a constant, finite number of signs (in particular it can be represented as a finite binary string).*

*Proof.*
- Let there be a m-non-compressible binary string of length $n$
- Algorithm $AS$ generates all strings of length $n$
- Algorithm $E$ will remove all m-compressible strings from the list
- Algorithm $C$ checks whether string $a$ was removed and prints 1 if it was and 0 otherwise
- Algorithm $AS + E + C$ realizes defined function for all strings longer than string encoding them

□

Therefore Chaitin-Gödel theorem is, in a general case, not true.

## 5 Remarks and conclusions

A general strategy of proving through the diagonal argument, relies on showing non-equinumerosity of sets. In particular we show, that there exists $C$ functions $f : \mathbb{N} \to \mathbb{N}$ and only $\aleph_0$ computable functions.

Let us observe the following:

- Each problem which is possible to express, can be represented with a binary string
- Each problem which is possible to express in a finite way, can be represented with a finite binary string
- If the problem has a solution, then it can be represented as a binary string (finite or not)
- There is at most as many problems possible to express, as there is binary strings
- There is the same amount of solutions possible to express
- We are able to generate all and arbitrary binary strings with absolute accuracy (i.e. arbitrary long)
- Therefore each problem, which has a solution possible to express, can be solved with an arbitrary precision, however not necessarily in a finite time (e.g. computing $\pi$ or veryfying some theorems with empirical mathematics are infinite procedures).
- Each machine, which is able to transform every string (data) into any other (result) is a fully universal machine, which is not limited. Operations which are necessary and sufficient are: flipping arbitrary bit of input data and ability to extend/truncate the input.

Formalisms or algorithms (programs) and machines which contain equivalent of $AK$ axiom and are capable of generating all binary strings, are not limited by limitative theorems of Gödel, Turing, Church and Chaitin. A strong interpretation of limitative theorems, i.e. interpretation which assumes limitation of all sufficiently rich systems ignores three kinds of facts:

- existance of finite algorithms capable of generating all possible strings with a *b*rute force strategy

- existance of random generators, capable of generating new information

- existance and meaning of "reflexivity blockers" such as confirming sentence $ap$ or tester $T'$.

The process of proving theorems, can be represented as a process of solving a decision problem. Assumptions / input data are axioms and the theorem to be prooven $(A, Th)$, both possible to express with binary strings. Proving contains of a certain amount of steps of a program $SPr$. At the end we obtain a logical value $W(Th)$ that is a stringth of length 1 (0 or 1). $(A, Th) \to SPr \to W(Th)$. The $(A, Th)$ string represents some amount of information, however $SPr$ can represent more (!) because order of system's syntax rules and way in which they are used can allow that. In particular $SPr$ does not have to be non-compressible to $(A, Th)$ and usually it is not. It means, that by proving theorems we generate new information - that is the essence of mathematical creativity and mathematical genius. System in which a certain theorem was proved is a different one, than one in which this was not done, beacause it represents more information.